\documentclass{iau} 
\usepackage{graphicx}
\usepackage{hyperref,natbib}
\usepackage{amsmath}
\usepackage{color}
\usepackage{ulem}
\definecolor{darkgreen}{rgb}{0.13, 0.55, 0.13}
\definecolor{brown}{rgb}{0.59, 0.29, 0.0}
\definecolor{ab}{rgb}{0.36, 0.54, 0.66}

\newcommand{\aref}[1]{\hyperref[#1]{Appendix~\ref{#1}}}

\title[Nitrogen enrichment at low metallicities] 
{Can pre-supernova winds from massive stars enrich the interstellar medium with nitrogen at high redshift?}

\author[A Roy et al.]   
{Arpita Roy$^1$,
Mark R. Krumholz$^{2,3}$,
Michael A. Dopita$^{2\thanks{deceased}}$,
Ralph S. Sutherland$^{2, 3}$,
Lisa J. Kewley$^{2, 3}$,
Alexander Heger$^{3, 4}$}

\affiliation{$^1$Scuola Normale Superiore, Piazza dei Cavalieri 7, 56126 Pisa, Italy \\ email: {\tt arpita.roy@sns.it}\\
$^{2}$RSAA, Australian National University, Cotter Road, Weston Creek, ACT 2611, Australia.\\
$^{3}$ASTRO 3D, Canberra, ACT 2611, Australia\\
$^{4}$School of Physics and Astronomy, Monash Centre for Astrophysics, 19 Rainforest walk, Monash University, VIC 3800, Australia.\\
}

\pubyear{2022}
\volume{366}  
\jname{The origin of outflows in evolved stars}
\editors{L. Decin, A. Zijlstra, C. Gielen, eds.}

\def\apj{\textit{ApJ}}               
\def\apjs{\textit{ApJS}}               
\def\apss{\textit{Ap\&SS}}             
\def\aap{\textit{A\&A}}                
\def\mnras{\textit{MNRAS}}

\begin{document}

\maketitle

\begin{abstract}
Understanding the nucleosynthetic origin of nitrogen and the evolution of the N/O ratio in the interstellar medium is crucial for a comprehensive picture of galaxy chemical evolution at high-redshift because most observational metallicity (O/H) estimates are implicitly dependent on the N/O ratio. The observed N/O at high-redshift shows an overall constancy with O/H, albeit with a large scatter. We show that these heretofore unexplained features can be explained by the pre-supernova wind yields from rotating massive stars (M$\gtrsim 10 \, \mathrm{M}_\odot$, $v/v_{\rm{crit}} \gtrsim 0.4$). Our models naturally produce the observed N/O plateau, as well as the scatter at low O/H. We find the scatter to arise from varying star formation efficiency. However, the models that have supernovae dominated yields produce a poor fit to the observed N/O at low O/H. This peculiar abundance pattern at low O/H suggests that dwarf galaxies are most likely to be devoid of SNe yields and are primarily enriched by pre-supernova wind abundances. 

\end{abstract}

\firstsection 
\section{Introduction}
\label{sec:intro}

The origin of nitrogen in the Universe and its cosmic evolution is a question at the forefront of current astrophysical studies. Observationally, the origin of N and its evolution play an important role in optical astronomers' metallicity determinations, which are heavily reliant on the N/O ratio. In fact, one of the best metallicity diagnostics is based on the [NII]/[OII] ratio \citep{kewley2002}. However, at high redshift the OII doublet is often unobservable, and therefore the metallicity estimate is dependent on either [NII]/H$\alpha$ \citep{denicolo2002} or [NII]/[OIII] \citep{pettini2004} ratio, and in the worst-case scenario when only the red lines of H$\alpha$, [NII]$\lambda$6584 and [SII]$\lambda\lambda$6717, 6731 doublets are observed, the O/H estimate needs to depend on either [NII]/H$\alpha$ or [NII]/[SII], or a carefully chosen combination of the two \citep{dopita2016}.

Our understanding of the N/O ratio is poor because of the complex chemical origin of N. Unlike oxygen, which has a primary origin, N has both primary and secondary origin. An element is defined as a primary element when it is produced entirely via the nucleosynthesis process inside a star without any initial seeds from elements that the star inherited from the interstellar medium (ISM) at birth. By this definition, O is a primary element because it is produced in stars undergoing triple-$\alpha$ reactions that does not rely on pre-existing C, N, or O that was imbibed by the star at birth. On the other hand, N can be produced during the core H burning in massive stars via CNO nucleosynthesis where the C used is inherited from the ISM, and hence this channel of N production is secondary. However, during core He burning in massive stars, shell H burning is catalysed by C produced inside the star itself, and hence this N production is primary. We discuss these two N production scenarios in detail in section \autoref{sec:Nitro}.

\section{Chemical origin of nitrogen and the associated dredge-up mechanisms}
\label{sec:Nitro}

A massive star has a large central convective zone where chemical elements are quasi-homogeneously mixed, its immediate outer layer is a radiative zone, and the outermost one is a thin convective zone. In rotating stars, various rotational instability induced diffusions of chemical elements dredge up the heavy elements from the core to the surface, as has been proposed by several authors in the past \citep{heger2000, meynet2000, maeder2005, meynet2005}. These rotational mixings behave as a bridge between the inner and the outer convective zones that helps in transporting the heavy elements from the core to the surface crossing the radiative barrier. However, in non-rotating stars, dredging up heavy metals from the inner convective zone to the outer convective layer crossing the radiative barrier becomes challenging. \cite{roy2020} proposes that the mechanism of the exposure of ``fossil"-convective cores,  where the star exposes regions that are no longer convective but were part of the convective core at an early stage in the evolution for a metal-rich ([$\mathrm{Fe}/\mathrm{H}$]$\geq -1.0$) massive ($\geqslant 80 \, \mathrm{M}_\odot$) star, can dredge up the heavy elements from the core to the surface. This happens because the star shrinks as it loses mass by main-sequence winds. They also find that even the modest amount of mass loss in these stars can expose the ``fossil"-convective cores, thereby enhancing the surface abundances of heavy elements.

Given the various dredge-up mechanisms, C produced in the He burning core, undergoing the triple-$\alpha$ process, gets dredged up to the surface and used as a catalyst to N production in the H burning shell. In this method, the H burning shell uses the C that is self-produced by the star for the CNO cycle to convert the C to N, and therefore this N production channel has a primary origin. Whereas, when the core is H burning undergoing CNO process, the C that is converted to N comes from the ISM at birth, and therefore that N has a secondary origin. As O has a primary origin, if N has a primary origin too, then we will expect N/O ratio to be constant with O/H. However, if N has a secondary origin, then we will see N/O to increase linearly with O/H. 

\section{Observations of N enrichment at high-redshift}
\label{sec:N_highZ}
Given the challenges and complexities of accurate modelling of N production, many authors aim to probe the origin of N by observing the N/O trend with O/H in metal-poor systems. There are three major approaches frequently undertaken for these observations. Firstly, one can observe metal-poor Galactic halo stars as they carry the footprints of early N production in our Galaxy. Secondly, one can use the HII regions of the Milky-Way and nearby galaxies at various global metallicities for the studies of N production as these systems behave as proxies of galaxies at different redshifts. Thirdly, one can observe high-redshift galaxies directly to study {\it in situ} N and O production.

\cite{israelian2004} follows the first approach, and they survey 31 unevolved halo stars within the metallicity range of $-3.1 \leq [\mathrm{Fe}/\mathrm{H}] \leq 0$. They find constancy of [N/O] below [O/H]$\leqslant -1.8$ hinting towards the primary origin of N, and a turnover beyond that metallicity with $[\mathrm{N}/\mathrm{O}] \propto [\mathrm{O}/\mathrm{H}]$ suggesting the secondary origin of N at high metallicity. In a similar spirit, \cite{spite2005} also analyse 35 stars with $-4 \leq [\mathrm{Fe}/\mathrm{H}] \leq -2$. However, contrary to \cite{israelian2004}'s observations, they find a broad scatter in [N/O] with no systemic trend in [N/O] with [O/H] at these low metallicities. This suggests a probable more complex origin of N rather than the simple primary origin of N as proposed before.

Observations of HII regions in our Galaxy and local dwarfs also suggest a similar complex origin for N. \cite{izotov1999} measure N and O yields by observing 50 blue compact galaxies with metallicity $\mathrm{Z}_\odot/50 < \mathrm{Z} < \mathrm{Z}_\odot/7$, and find that N/O is constant with O/H at low metallicity. However, contrary to these observations, several other authors find a large scatter of $\sim 1$ dex in N/O for a given O/H at low metallicity \citep{kobulnicky1996, liang2006, perez2005, perez2009} suggesting no systematic trend in N/O, similar to the observations of metal-poor halo stars.

Combining both these observations from stars and gas in the HII regions, we conclude that there is an overall plateau in N/O ($\log(\mathrm{N}/\mathrm{O}) \sim -1.7$) at low metallicity ($12+\log(\mathrm{O}/\mathrm{H}) \leqslant 7.5$), albeit with a large scatter ($\sim 1$ dex).

\section{Stellar models and adundance calculation}
\label{sec:abund}

Having discussed various observational approaches to measure N and O abundances at low metallicities, we discuss our theoretical models to estimate yields from winds and supernovae (SNe) in this section. For stellar models, we use modified MESA Isochrone Stellar Tracks II (MIST-II, Dotter et al., in prep.) that are appropriate for the evolution of massive stars. In this modified version, we include a realistic treatment for diffusion and transport of chemical elements driven by various rotational instabilities and also more accurate implementation of $\alpha$ enhancement to emulate Galactic concordance abundances. For details, we refer readers to \cite{roy2020, roy2021, grasha2021}. In this proceeding, we show results of our stellar models with masses $10 \, \mathrm{M}_\odot \leq \mathrm{M} \leq 150 \, \mathrm{M}_\odot$ for our fiducial rotation rate (see \cite{roy2021}) of $v/v_{\rm{crit}}=0.4$ for three metallicities, $[\mathrm{Fe}/\mathrm{H}]=-2.0$, $-3.0$, and $-4.0$. For SN models, we adopt the yields from \cite{limongi2018}.

For the yield estimate, we consider a simple stellar population with the initial mass function (IMF) $\mathrm{d}n/\mathrm{d}m$ at the time of formation $t=0$. We use the Salpeter IMF with the upper mass limit of $150 \, \mathrm{M}_\odot$. The total mass of an element $X$ per unit stellar mass (at formation) that is returned to the ISM gas phase at time t is

\begin{eqnarray}
\lefteqn{\psi_{\rm ret}(X,t) = \psi_w(X,t) + \psi_{\rm SN}(X,t) =}
\label{eq:psi_Xt}
\\
& & \frac{1}{\langle m \rangle} \left[
\int_0^\infty M_{w}(X,m,t) \frac{{\mathrm{d}}n}{{\mathrm{d}}m} \, {\mathrm{d}}m +
\int_{m_{\rm d}}^\infty M_{\rm SN}(X,m) \frac{{\mathrm{d}}n}{{\mathrm{d}}m} \, {\mathrm{d}}m\right],
\nonumber
\end{eqnarray}
where $\psi_w$, $\psi_{\rm SN}$ are wind and SN contribution respectively, 
\begin{equation}
\langle m\rangle = \int m\frac{dn}{dm}\,dm
\end{equation}
is the mean stellar mass, $m_{\rm d}$ is the ``death mass'' at $t$, and, 
\begin{equation}
M_w(X,m,t) = \int_0^t \dot{M}_w(X,m,t')\, dt'
\end{equation}
is the wind ejected cumulative mass of an element $X$ by a star of initial mass $m$ up to age $t$.

Now we assume a gas reservoir of mass $M_\mathrm{g}$ that converts $\epsilon_*$ of its mass instantaneously to stars. If the mass fraction of an element $X$ prior to this star-formation event is $f_0(X)$, then at a time $t$, after the star formation, the mass of that element in the gas phase is 
\begin{equation}
M(X,t) = \left[(1 - \epsilon_*) f_0(X) + \epsilon_* \psi_{\rm ret}(X,t)\right] M_{\rm g}.
\label{eq:M_Xt}
\end{equation}
We can therefore write the mass ratio of two elements at time $t$ as,
\begin{equation}
\frac{M(X,t)}{M(Y,t)} = \frac{(1 - \epsilon_*) f_{0}(X) + \epsilon_* \psi_{\rm ret}(X,t)}{(1 - \epsilon_*) f_{0}(Y) + \epsilon_* \psi_{\rm ret}(Y,t)}\, .
\label{eq:Xi_Xj}
\end{equation}
We also consider that subsequent to the star formation, the galaxy accretes an additional primordial pristine gas of mass $M_{\rm{p}}$ with the fractional abundance of element $X$ as $f_{\mathrm{p}}(X)$. In this scenario, the mass ratio of two elements becomes
\begin{equation}
\frac{M(X,t)}{M(Y,t)} = \frac{(1 - \epsilon_*) f_{0}(X) + \epsilon_* \psi_{\rm ret}(X,t) + \epsilon_{\rm p}(t) f_{\rm p}(X)}{(1 - \epsilon_*) f_{0}(Y) + \epsilon_* \psi_{\rm ret}(Y,t) + \epsilon_{\rm p}(t) f_{\rm p}(Y)} \, ,
\label{eq:Xi_Xj_pri}
\end{equation}
where $\epsilon_{\rm p}(t) = M_{\rm p}(t) / M_{\rm g}$ is the fractional mass of the added primordial gas. This abundance ratio in terms of number fractions can be rewritten as:
\begin{equation}
\frac{N(X,t)}{N(Y,t)} = \frac{M(X,t)}{M(Y,t)} \frac{m_Y}{m_X} \, ,
\label{eq:Ni_Nj}
\end{equation}
where $m_X$ is the atomic mass of element $X$. We denote $N(X)/N(Y)$ as $X/Y$. 

\begin{figure*}
\centerline{
\includegraphics[width=1.0\textwidth]{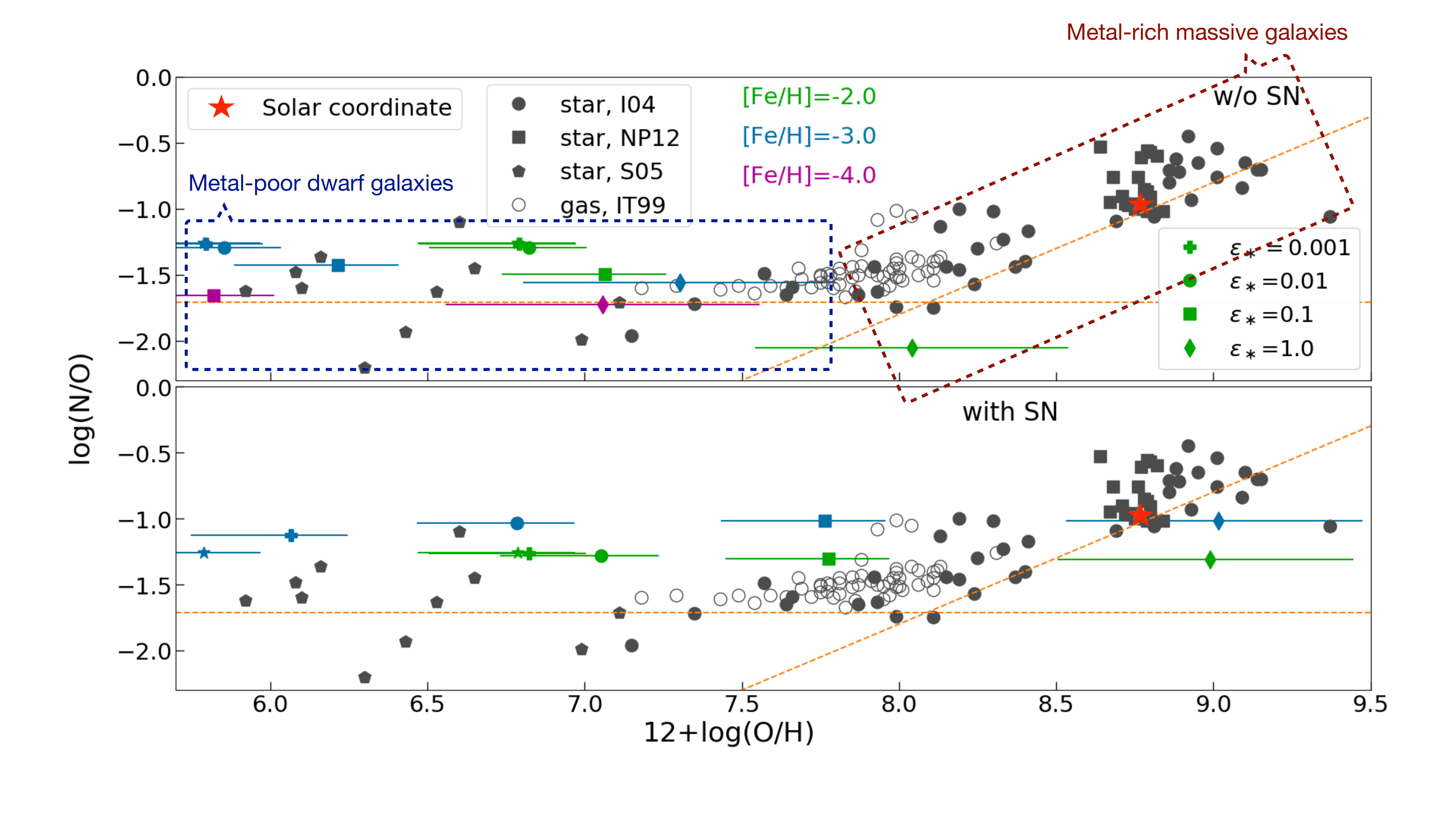}
}
\caption{Adapted from Figure 8 of \cite{roy2021}: $\log$(N/O) vs.~$\log$(O/H) from our models (coloured data points) and observations (black data points). {\it Top panel:} Models of galaxies that retain only the low-velocity pre-SN wind yields, {\it Bottom panel:} galaxies that retain both SNe ejecta and pre-SN winds. Black filled points (representing stellar yields) show the observations of several authors: \cite{israelian2004} (I04, circles),  \cite{spite2005} (S05, pentagons),  \cite{nieva2012} (NP12, squares), and open circles represent the HII regions (IT99, \cite{izotov1999}). The orange dashed lines show the primary and secondary N production channels as proposed by \cite{dopita2016}. The red star represents the solar value. The dashed blue and brown boxes divide the N/O versus O/H phase space into two regimes based on our theoretical predictions that metal-poor dwarfs (blue box) retain only wind ejecta, leading to a low N/O ratio with large scatter characteristic of wind yields, while metal-rich massive galaxies (brown box) retain both wind and SN ejecta, leading to a higher N/O ratio characteristic of SNe. 
Coloured points represent theoretical predictions for different initial metallicity $[\mathrm{Fe}/\mathrm{H}]$ and star formation efficiency $\epsilon_*$ as indicated by the figure legend. For each point, the horizontal bar corresponds to the scenario where a galaxy accretes an additional fraction of primordial ($\epsilon_{\rm p}$) hydrogen and helium followed by the star-formation event. Our assumed range of $\epsilon_{\rm p}$ are $\log \epsilon_{\rm p} = -0.5 - 0.5$. Note that, we omit the $[\mathrm{Fe}/\mathrm{H}]=-4$ case for SN yields because our SN yield tables only go down to $[\mathrm{Fe}/\mathrm{H}]=-3$ as adopted from \cite{limongi2018}.
}
\label{fig:NO_OH}
\end{figure*}

\section{N/O and O/H distributions}
\label{sec:NO_OH}
Having discussed our model yield calculation in \autoref{sec:abund}, we now discuss the distribution of N/O versus O/H to predict the observed N/O ratio trend at low metallicity. We show our predicted N/O$\textendash$O/H distribution for different parameter combinations and for our fiducial rotation rate of $v/v_{\rm{crit}}=0.4$ (for justification, see \cite{roy2021}) for three metallicities, $[\mathrm{Fe}/\mathrm{H}]=-2$, $-3$, and $-4.0$ in \autoref{fig:NO_OH}. We find that when galaxies only retain the wind yields, we see a plateau in N/O with $\log(\mathrm{N}/\mathrm{O}) \sim -1.5 \, \textendash \, -2.0$ with a scatter of $\sim 1$ dex at low metallicities. The scatter in N/O readily arises from the variation in metallicities and star formation efficiencies. This prediction of the plateau and scatter strikingly matches with the observed features as discussed in \autoref{sec:N_highZ}. This result indicates that the metal-poor dwarf galaxies might have their metal abundances solely from wind yields of massive stars, with high-velocity SNe ejecta escaping them given their shallow potential wells. On the other hand, we also notice that galaxies that can retain both wind and SNe ejecta have higher values of N/O with $\log(\mathrm{N}/\mathrm{O}) \sim -1$. This suggests that massive metal-rich galaxies are more suitable candidates that can retain both these low and high-velocity ejecta, and, therefore might have high values of N/O. Also, this high N/O ratio matches well with the observed N/O upturn at high O/H  ($12+\log(\mathrm{O}/\mathrm{H})\gtrsim 8.0$). Thus, this observed upturn in N/O might partially be coming from SNe contribution in metal-rich massive galaxies along with the onset of secondary nitrogen production in AGB stars.

Our modelled yields depend on two major parameters $\epsilon_*$ and $\epsilon_{\rm p}$, other than stellar metallicity. The dependence on $\epsilon_{\rm p}$ can be easily comprehended. Changing $\epsilon_{\rm p}$, the parameter that controls the amount of primordial gas added to the galaxy, does not vary N/O ratio, it only changes the O/H ratio, and therefore slides points horizontally in \autoref{fig:NO_OH}. Thus, the more pristine primordial gas is accreted by the high-redshift galaxies, the metal-poor they become. Understanding the dependence of N/O versus O/H on $\epsilon_*$ is more subtle. As the star formation efficiency increases, the O/H ratio increases, whereas the N/O ratio decreases. O/H increases because the more the star-formation efficiency, the more the astrated material gets deposited into the ISM thereby enhancing the gas phase O abundance. However, to understand the decrease in the N/O ratio with increasing $\epsilon_*$, we will first have to understand the pure wind-driven case ($\epsilon_*=1$) without any contribution from the pre-existing ISM abundance. For this case, as the majority of massive stars enter the core He-burning phase, primary O is produced more in amount in the triple-$\alpha$ process compared to the primary N production in the shell H-burning. This causes the reduction in the N/O ratio. Having said that, as we decrease $\epsilon_*$, the less amount of stellar astrated yield is deposited into the ISM, and the more of the pre-existing ISM gas abundance start playing a stronger role. We find that the ISM N/O ratio prior to the star formation event is higher compared to the wind ejected N/O. Therefore, as we decrease $\epsilon_*$, the N/O ratio increases.

Combining these findings together, we conclude that we predict a plateau in N/O at low O/H, albeit with a large scatter, and this prediction matches pretty well with observations of metal-poor halo stars and local dwarfs. We also propose that this trend in N/O at low O/H is a signature that metal-poor dwarf galaxies may retain only the low-velocity wind yields, and the high-velocity SNe ejecta will most likely escape their low gravitational potential wells. On the other hand, metal-rich massive galaxies are likely to retain both wind and SNe ejecta, thereby having a significantly higher N/O ratio, and therefore they might be major contributors to the upturn in N/O at high O/H ($12+\log(\mathrm{O}/\mathrm{H})\gtrsim 8.0$) along with the secondary N from AGB stars.

\section{Conclusions}
We conclude this paper with three salient findings:
\begin{itemize}
\item Winds from metal-poor ($[\mathrm{Fe}/\mathrm{H}] \leq -2.0$) massive ($\geq 10 \, \mathrm{M}_\odot$) stars produce substantial nitrogen with the N/O ratio of $\log(\mathrm{N}/\mathrm{O}) \sim -2 \textendash -1.5$, independent of total oxygen metallicity. Dwarf galaxies are more likely to retain these low-velocity wind yields, and therefore this predicted N/O ratio is commonly observed in local dwarfs, along with metal-poor halo stars. 
\item In addition to the mean N/O ratio produced in winds providing a good match to the mean observed values, we also notice the scatter in N/O, similar to observations. In our modelling, this scatter arises from varying metallicity and star formation efficiency.
\item SNe produce a higher N/O ratio with $\log(\mathrm{N}/\mathrm{O}) \sim -1$ compared to winds. Also, the high-velocity SNe ejecta are more likely to be retained by more massive, metal-rich galaxies. Therefore, SNe ejecta might play an important role in the upturn in N/O ratio at higher O/H ($12+\log(\mathrm{O}/\mathrm{H}) \gtrsim 8$) along with the secondary N produced in AGB stars.
\end{itemize}

\label{sec:conc}

\end{document}